\documentclass[12pt]{article}

\usepackage{graphicx}

\title{A Simple Classification of Solitons}
\author{ Yousef Yousefi and  Khikmat Kh. Muminov\\
Physical-Technical Institute named after S.U.Umarov\\
 Academy of Sciences of Republic of Tajikistan\\
Aini Ave 299/1, Dushanbe, Tajikistan}
\date{}
\begin{document}
\maketitle

\begin{abstract}
In this report, fundamental educational concepts of linear and non-linear equations and solutions of nonlinear equations from the book High-Temperature Superconductivity: The Nonlinear Mechanism and Tunneling Measurements (Kluwer Academic Publishers, Dordrecht, 2002, pages 101-142) is given. There are a few ways to classify solitons. For example,  there are topological and nontopological solitons. Independently of the topological nature of solitons, all solitons can be divided into two groups by taking into account their profiles: permanent and timedependent. For example, kink solitons have a permanent profile (in ideal systems), while all breathers have an internal dynamics, even, if they are static. So, their shape oscillates in time. The third way to classify the solitons is in accordance with nonlinear equations which describe their evolution. Here we discuss common properties of solitons on the basis of the four classification.

\end{abstract}

\section{Introduction}
For a long time linear equations have been used for describing different phenomena. For example, Newton’s, Maxwell’s and Schrödinger’s equations are linear, and they take into account only a linear response of a system to an external disturbance. However, the majority of real systems are nonlinear. Most of the theoretical models are still relying on a linear description, corrected as much as possible for nonlinearities which are treated as small perturbations. It is well known that such an approach can be absolutely wrong. The linear approach can sometimes miss completely some essential behaviors of the system. 

Nonlinearity has to do with thresholds, with multistability, with hysteresis, with phenomena which are changed qualitatively as the excitations are changed. In a linear system, the ultimate effect of the combined action of two different causes is merely the superposition of the effects of each cause taken individually. But in a nonlinear system adding two elementary actions to one another can induce dramatic new effects reflecting the onset of cooperativity between the constituent elements. To understand nonlinearity, one should first understand linearity. Consider linear waves. In general, a wave may be defined as a progression through matter of a state of motion. Characteristic properties of any linear wave are:

 (i) the shape and velocity of a linear wave are independent of its amplitude;

 (ii) the sum of two linear waves is also a linear wave; and

 (iii) small amplitude waves are linear. Fig.1a shows an example of a periodic linear wave.

 Large amplitude waves may become nonlinear. The fate of a wave travelling in a medium is determined by properties of the medium. Nonlinearity results in the distortion of the shape of large amplitude waves, for example, in turbulence. However, there is another source of distortion—the dispersion of a wave.

More than 100 years ago the mathematical equations describing solitary waves were solved, at which point it was recognized that the solitary wave, shown in Fig.1b, may exist due to a precise balance between the effects of nonlinearity and dispersion. Nonlinearity tends to make the hill steeper (see Fig. 1b), while dispersion flattens it.
 
\begin{figure}
\centering
\includegraphics[width=4 in]{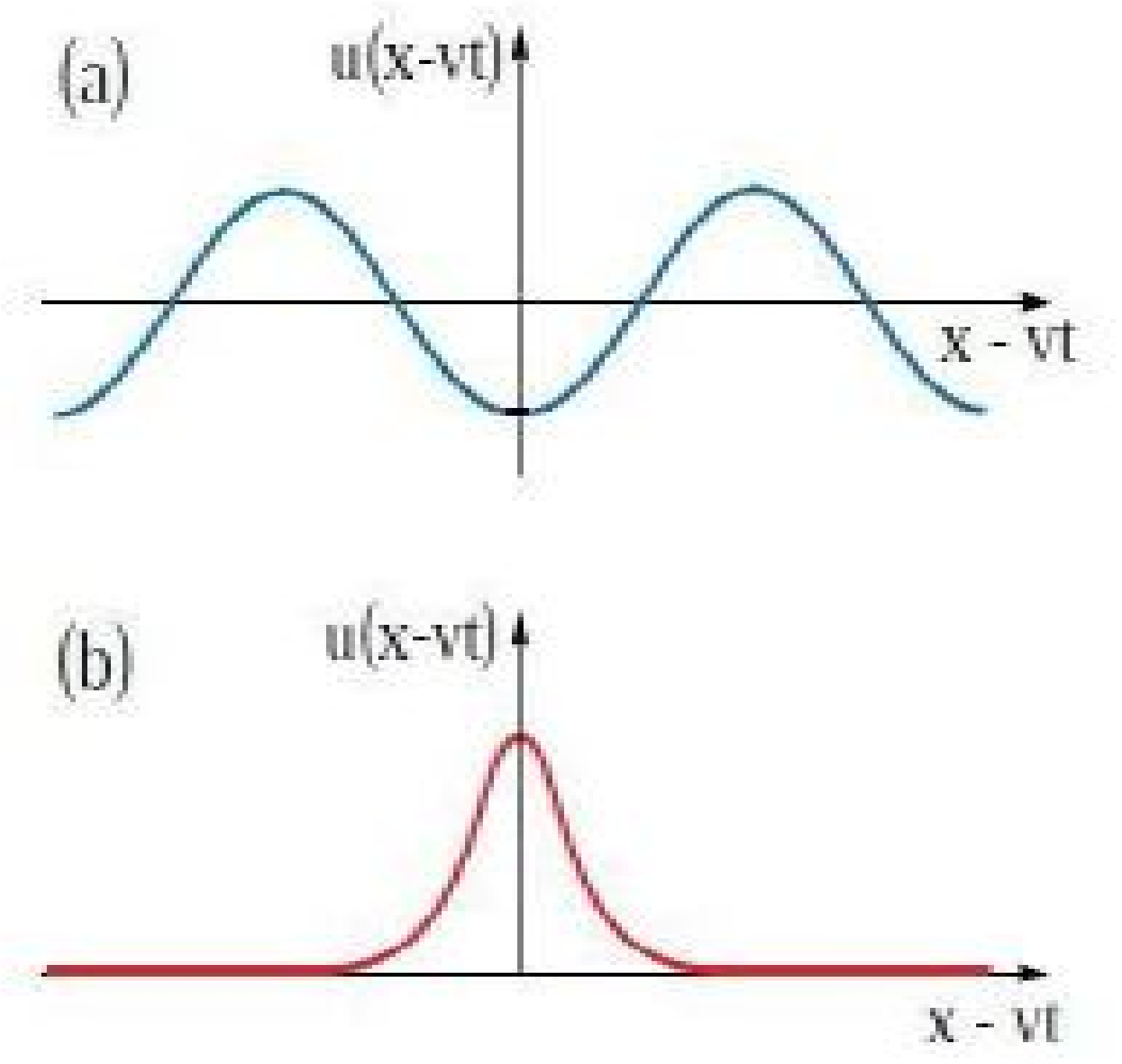}
\caption{Sketch of (a) a periodic linear wave, and (b) a solitary wave.}
\end{figure}

The solitary wave lives “between” these two dangerous, destructive “forces.” Thus, the balance between nonlinearity and dispersion is responsible for the existence of the solitary waves. As a consequence, the solitary waves are extremely robust. Solitary waves or solitons cannot be described by using linear equations. Unlike ordinary waves which represent a spatial periodical repetition of elevations and hollows on a water surface, or condensations and rarefactions of a density, or deviations from a mean value of various physical quantities, solitons are single elevations, such as thickenings etc., which propagate as a unique entity with a given velocity. The transformation and motion of solitons are described by nonlinear equations of mathematical physics. The history of solitary waves or solitons is unique. The first scientific observation of the solitary wave was made by Russell in 1834 on the water surface[1]. One of the first mathematical equations describing solitary waves was formulated in 1895. And only in 1965 were solitary waves fully understood! Moreover, many phenomena which were well known before 1965 turned out to be solitons! Only after 1965 was it realized that solitary waves on the water surface, nerve pulse, vortices, tornados and many others belong to the same category: they are all solitons! That is not all, the most striking property of solitons is that they behave like particles!. Other important properties of soliton are:

1. It does not change shape.

2. In a region of space is limited.

3. After dealing with other solitons, keep its shape.

 Mathematically, there is a difference between “solitons” and “solitary waves.” Solitons are localized solutions of integrable equations, while solitary waves are localized solutions of non-integrable equations. Another characteristic feature of solitons is that they are solitary waves that are not deformed after collision with other solitons. Thus the variety of solitary waves is much wider than the variety of the “true” solitons. Some solitary waves, for example, vortices and tornados are hard to consider as waves. For this reason, they are sometimes called soliton-like excitations. To avoid this bulky expression we shall often use the term soliton in all cases.

\section{Classification of solitons}

There are a few ways to classify solitons[33]. For example, as we known, there are topological and nontopological solitons. Independently of the topological nature of solitons, all solitons can be divided into two groups by taking into account their profiles: permanent and timedependent. For example, kink solitons have a permanent profile (in ideal systems), while all breathers have an internal dynamics, even, if they are static. So, their shape oscillates in time. The third way to classify the solitons is in accordance with nonlinear equations which describe their evolution. Here we discuss common properties of solitons on the basis of the four classification.

\subsection{Classical and quantum solitons}

A rough description of a classical soliton is that of a solitary wave which shows great stability in collision with other solitary waves. A solitary wave, as we have seen, does not change its shape, it is a disturbance $ u(x-ct)$  which translating along the x-axis with speed c. [2]

A remarkable example for this type is soliton solution for linear dispersion less equation or KdV equation.
 
\begin{figure}
\centering
\includegraphics[width=2.5in]{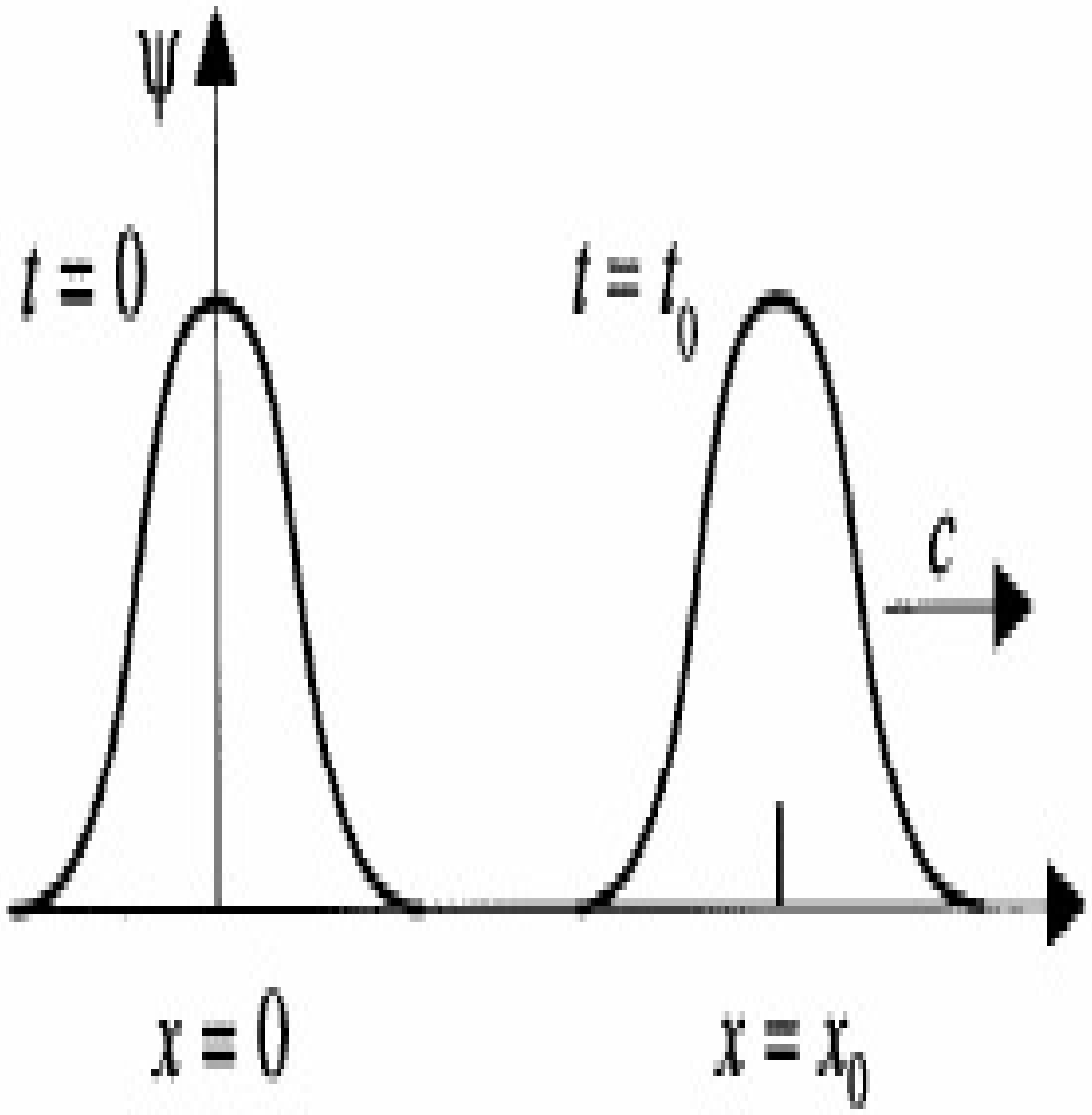}
\caption{Classical soliton.}
\end{figure}

Quantum solitons for physical systems governed by quantum attractive nonlinear Schrödinger model and quantum Sine-Gordon model. These solitons are coherent states or eigenvalues of annihilation operator $ \hat a $.

The one-dimensional quantum NLS equation, in term of quantum fields $\hat\psi(x,t),\hat \psi^+(x,t)$  is

\begin{eqnarray}
+i\bar h \frac{\partial\hat \psi}{\partial t}&=&-\frac{\bar h^2}{2m}\frac{\partial^2\hat \psi}{\partial x^2}+2c\hat \psi^+\hat \psi^2 \nonumber\\
-i\bar h  \frac{\partial \hat \psi^+}{\partial t}&=&-\frac{\bar h^2}{2m}\frac{\partial^2\hat \psi^+}{\partial x^2}+2c(\hat \psi^+)^2\hat \psi
\end{eqnarray}

Also for second model, Sine-Gordon model, equation is

\begin{eqnarray}
\frac{\partial^2 \hat \phi}{\partial x^2}-\frac{1}{c^2}\frac{\partial^2 \hat \phi}{\partial t^2}=m^2\bar h^{-2}c^4\hat \phi
\end{eqnarray}

\subsection{topological and non-topological solitons}

In renormalize relativistic local field theories all solitary waves are either non-topological or topological [3,4].

In non-topologically soliton, for example the water canal solitary solution to the KdV equation means that the boundary conditions at infinity are topologically the same for the vacuum as for the soliton. The vacuum can be non-degenerate but an additive conservation law is required.

But topologically soliton need a degenerate vacuum. The boundary conditions at infinity are topologically different for the solitary wave than for a physical vacuum state. The solitary of topological soliton is due to the distinct classes of vacuum at the boundaries where these boundary conditions are characterized by a particular correspondence (mapping) between the group space and coordinate space, and because these mappings are not continuously deformable into one another they are topologically distinct.

In mathematics and physics, a topological soliton or a topological defect is a solution of a system of partial differential equations or of a quantum field theory homotopically distinct from the vacuum solution; it can be proven to exist because the boundary conditions entail the existence of homotopically distinct solutions. Typically, this occurs because the boundary on which the boundary conditions are specified has a non-trivial homotopy group which is preserved in differential equations; the solutions to the differential equations are then topologically distinct, and are classified by their homotopy class. Topological defects are not only stable against small perturbations, but cannot decay or be undone or be de-tangled, precisely because there is no continuous transformation that will map them (homotopically) to a uniform or "trivial" solution.

Various different types of topological defects are possible, with the type of defect formed being determined by the symmetry properties of the matter and the nature of the phase transition. They include:

{\bf Domain walls}, two-dimensional membranes that form when a discrete symmetry is broken at a phase transition. These walls resemble the walls of closed-cell foam, dividing the universe into discrete cells.

{\bf Cosmic strings} are one-dimensional lines that form when an axial or cylindrical symmetry is broken.

{\bf Monopoles}, point-like defects that form when a spherical symmetry is broken, are predicted to have magnetic charge, either north or south (and so are commonly called "magnetic monopoles").

{\bf Textures} form when larger, more complicated symmetry groups are completely broken. They are not as localized as the other defects, and are unstable. Other more complex hybrids of these defect types are also possible.

Topological defects, of the cosmological type, are extremely high-energy phenomena and are likely impossible to produce in artificial Earth-bound physics experiments, but topological defects that formed during the universe's formation could theoretically be observed.

No topological defects of any type have yet been observed by astronomers, however, and certain types are not compatible with current observations; in particular, if domain walls and monopoles were present in the observable universe, they would result in significant deviations from what astronomers can see. Theories that predict the formation of these structures within the observable universe can therefore be largely ruled out.

In condensed matter physics, the theory of homotopy groups provides a natural setting for description and classification of defects in ordered systems. Topological methods have been used in several problems of condensed matter theory. Poénaru and Toulouse used topological methods to obtain a condition for line (string) defects in liquid crystals can cross each other without entanglement. It was a non-trivial application of topology that first led to the discovery of peculiar hydrodynamic behavior in the A-phase of superfluid Helium-3.

Unlike in cosmology and field theory, topological defects in condensed matter can be experimentally observed. Ferromagnetic materials have regions of magnetic alignment separated by domain walls. Nematic and bi-axial nematic liquid crystals display a variety of defects including monopoles, strings, textures etc. Defects can also been found in biochemistry, notably in the process of protein folding.

In quantum field theory, a non-topological soliton (NTS) is a field configuration possessing, contrary to a topological one, a conserved Noether charge and stable against transformation into usual particles of this field for the following reason. For fixed
charge Q, the mass sum of Q free particles exceeds the energy (mass) of the NTS so that the latter is energetically favorable to exist.

The interior region of an NTS is occupied by vacuum different from surrounding one. Thus a surface of the NTS represents a domain wall, which also appears as a topological defect in field theories with broken discrete symmetry. If infinite, the domain walls cause contradiction with cosmology. But the surface of an NTS is a closed finite wall so, if it exists in the Universe, it does not cause those contradictions. Another point is that if the topological domain wall is closed, it shrinks because of wall tension. As for the NTS surface, it does not shrink since the decreasing of the NTS volume would increase its energy.

Quantum field theory has been developed to describe the elementary particles. However in the middle seventieth it was found out that this theory predicts one more class of stable compact objects: non-topological solitons. The NTS represents an unusual coherent state of matter, called also bulk matter. Models were suggested for the NTS to exist in forms of stars, quasars, the Dark matter and nuclear matter.

An NTS configuration is the lowest energy solution of classical equations of motion possessing a spherical symmetry. Such a solution has been found for a rich variety of field Lagrangians. One can associate the conserved charge with global, local, Abelian and non-Abelian symmetry. It appears to be possible the NTS configuration with bosons as well as with fermions to exist. In different models either one and the same field carries the charge and binds the NTS, or there are two different fields: charge carrier an binding field.

The spatial size of the NTS configuration may be elementary small or astronomically large: depending on a model, i.e. the model fields and constants. The NTS size could increase with its energy until the gravitation complicates its behavior and finally causes the collapse. Although in some models the NTS charge is bounded by the stability (or metastability).

\subsection{Classification of solitons in bases of shape}
\subsubsection{bell soliton}
the soliton solution of KdV equation have a bell shape and a low frequency solitons. This soliton referred to as non-topological solitons.

\begin{figure}
\centering
\includegraphics[width=4 in]{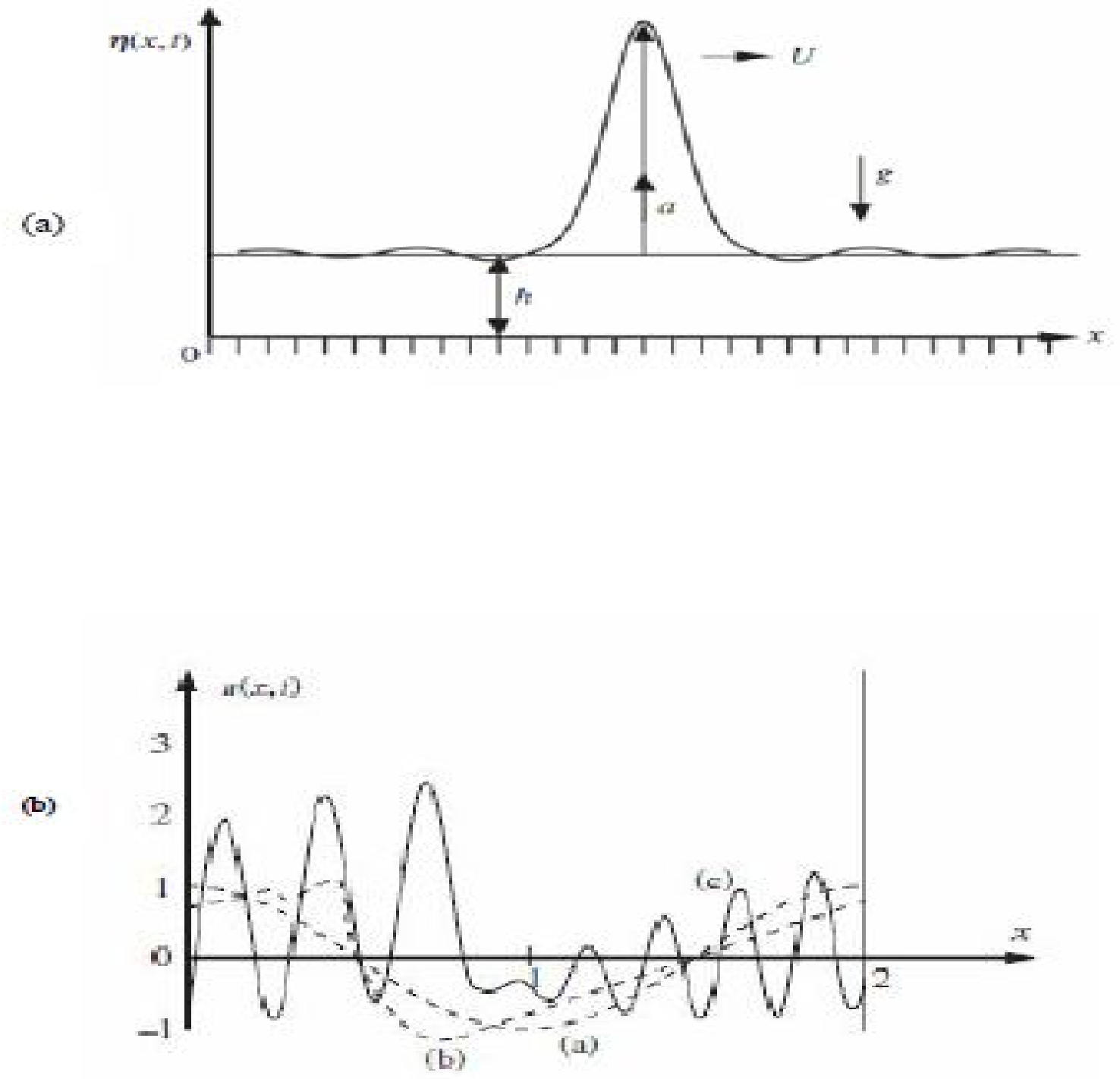}
\caption{ bell soliton, (a): solution of Kdv equation , (b): solution of HLS equation }
\end{figure}

the soliton solution of NLS equation have a bell shaped hyperbolic secant envelope modulated a harmonic (cosine) wave. This solution does not depend on the amplitude and high frequency soliton.

\subsubsection{Kink soliton}

The solutions of SC equation are called kink or anti-kink solitons, and velocity does not depend on the wave amplitude. This soliton referred to as topological solitons.

\begin{figure}
\centering
\includegraphics[width=4 in]{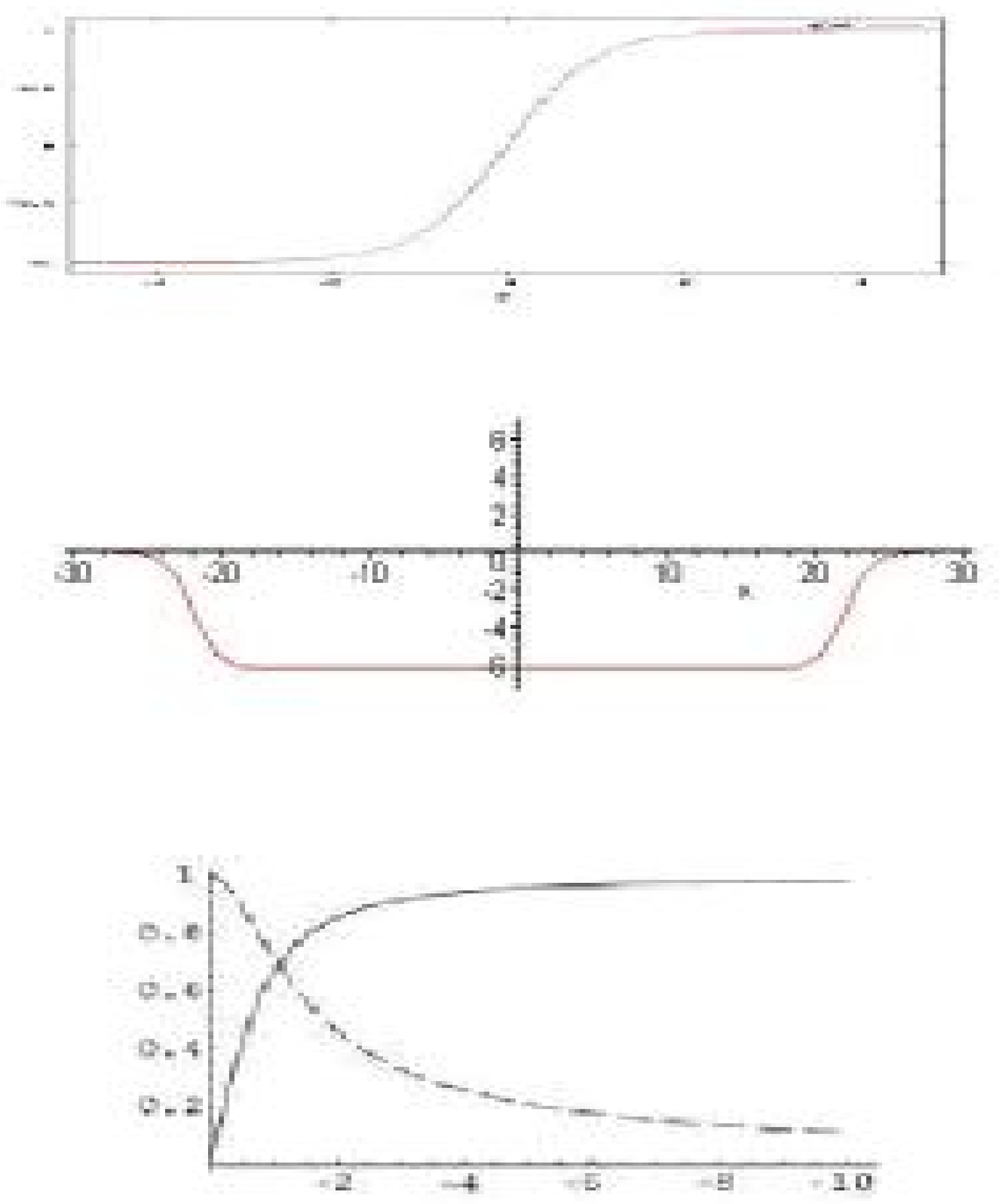}
\caption{Kink-antikink soliton solutions to the Sine-Gordon equation}
\end{figure}

\begin{figure}
\centering
\includegraphics[width=4 in]{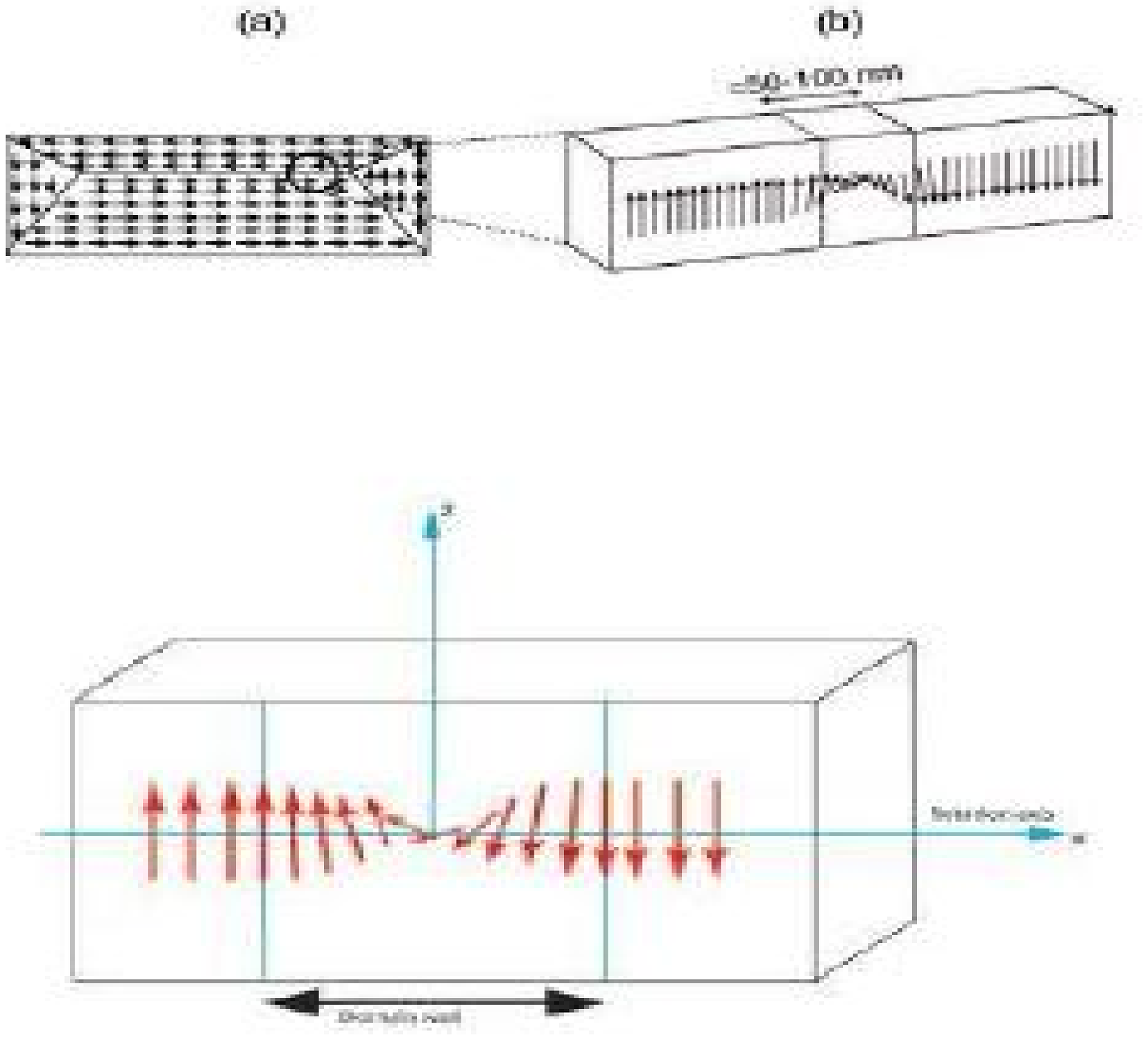}
\caption{Bloch wall between two ferromagnetic domains}
\end{figure}

A good physical example of a kink solution is a Bloch wall between two magnetic domains in a ferromagnet. The magnetic spins rotate from say, spin down in one domain to spin up in the adjacent domain. The transition region between down and up is called the Bloch wall. Under the influence of an applied magnetic field, the Bloch wall can propagate according to the Sine-Gordon equation.

\subsubsection{breather soliton}

Discrete breathers (DB), also known as intrinsic localized modes, or nonlinear localized excitations, are an important new phenomenon in physics, with potential applications of sufficient significance to rival or surpass the Soliton of integrable partial differential equations[6].  They occur in networks (includes all crystalline lattices and also quasicrystal and amorphous arrays) of oscillators (includes rotors and spins) rather than spatially continuous media, and are time-periodic spatially localized solutions.

\begin{figure}
\centering
\includegraphics[width=4 in]{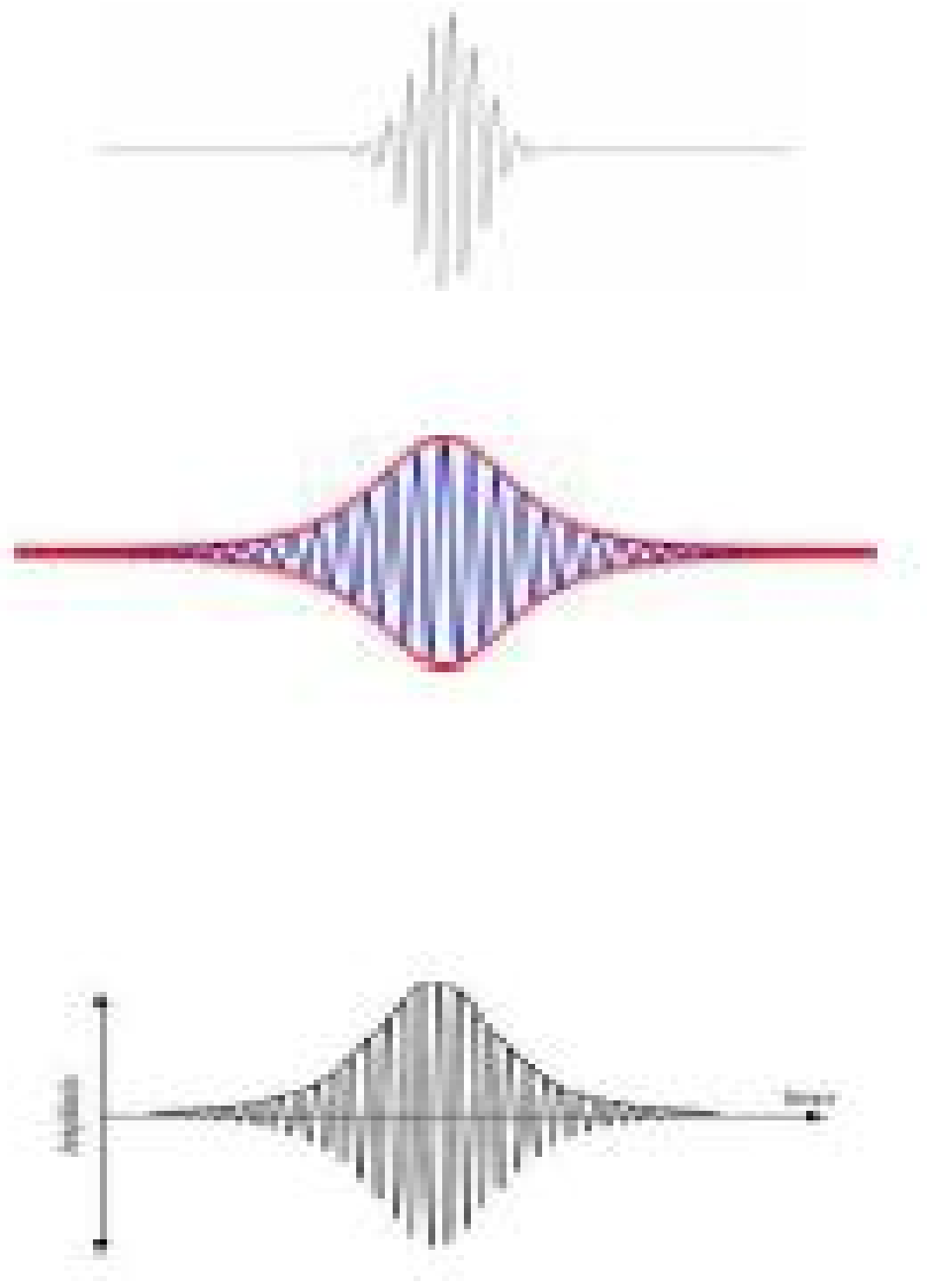}
\caption{breathers soliton}
\end{figure}

\subsection{Classification of solitons in bases of nonlinear equations}

Up to now we have considered two nonlinear equations which are used to describe soliton solutions: the KdV equation and the sine-Gordon equation. There is the third equation which exhibits true solitons it is called the nonlinear Schr¨odinger (NLS) equation[2]. We now summarize soliton properties on the basis of these three equations, namely, the Korteweg-de Vries equation:

\begin{eqnarray}
u_t=6uu_x-u_{xxx};
\end{eqnarray}

the sine-Gordon equation:

\begin{eqnarray}
u_{tt}=u_{xx}-sinu;
\end{eqnarray}

and the nonlinear Schrodinger equation:

\begin{eqnarray}
iu_t=-u_{xx}\pm |u|^2u;
\end{eqnarray}

where $ u_z $ means $ \frac{\partial u}{\partial z}$. For simplicity, the equations are written for the dimensionless function u depending on the dimensionless time and space variables.

There are many other nonlinear equations (i.e. the Boussinesq equation) which can be used for evaluating solitary waves, however, these three equations are particularly important for physical applications. They exhibit the most famous solitons: the KdV (pulse) solitons, the sine-Gordon (topological) solitons and the envelope (or NLS) solitons. All the solitons are one-dimensional (or quasi-one-dimensional). Figure .7 schematically shows these three types of solitons. Let us summarize common features and individual differences of the three most important solitons.

\begin{figure}
\centering
\includegraphics[width=4 in]{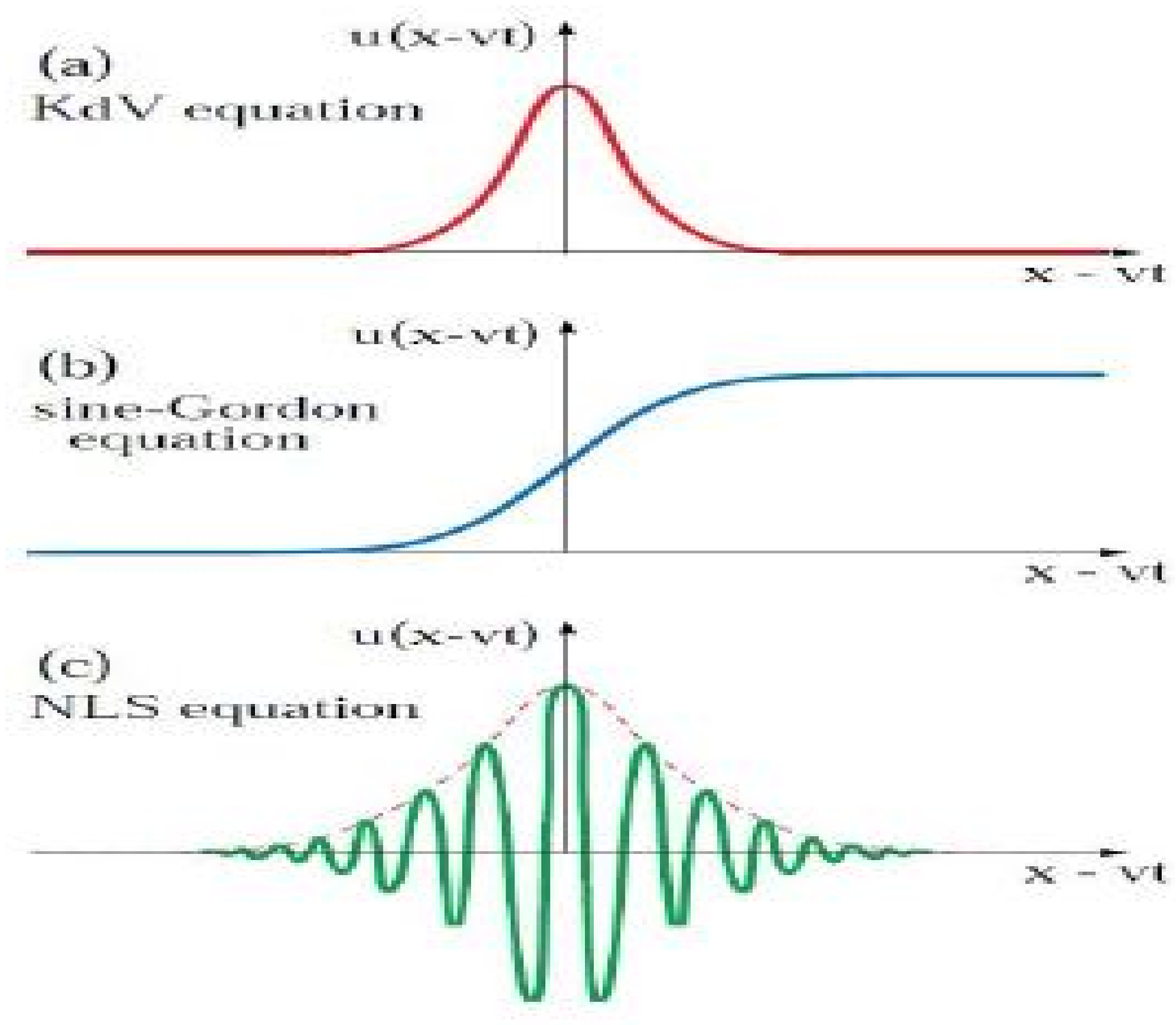}
\caption{Schematic of the soliton solutions of: (a) the Korteweg-de Vries equation; (b) the Sine-Gordon equation, and (c) the nonlinear schrodinger equation.}
\end{figure}

{\bf A. The KdV solitons}

The exact solution of the KdV equation is given by Eq. (7). The basic properties of the KdV soliton, shown in Fig. 7a, can be summarized as follows [7]:

i. Its amplitude increases with its velocity (and vice versa). Thus, they cannot exist at rest.

ii. Its width is inversely proportional to the square root of its velocity.

iii. It is a unidirectional wave pulse, i.e. its velocity cannot be negative for solutions of the KdV equation.

iv. The sign of the soliton solution depends on the sign of the nonlinear coefficient in the KdV equation.

\begin{figure}
\centering
\includegraphics[width=4 in]{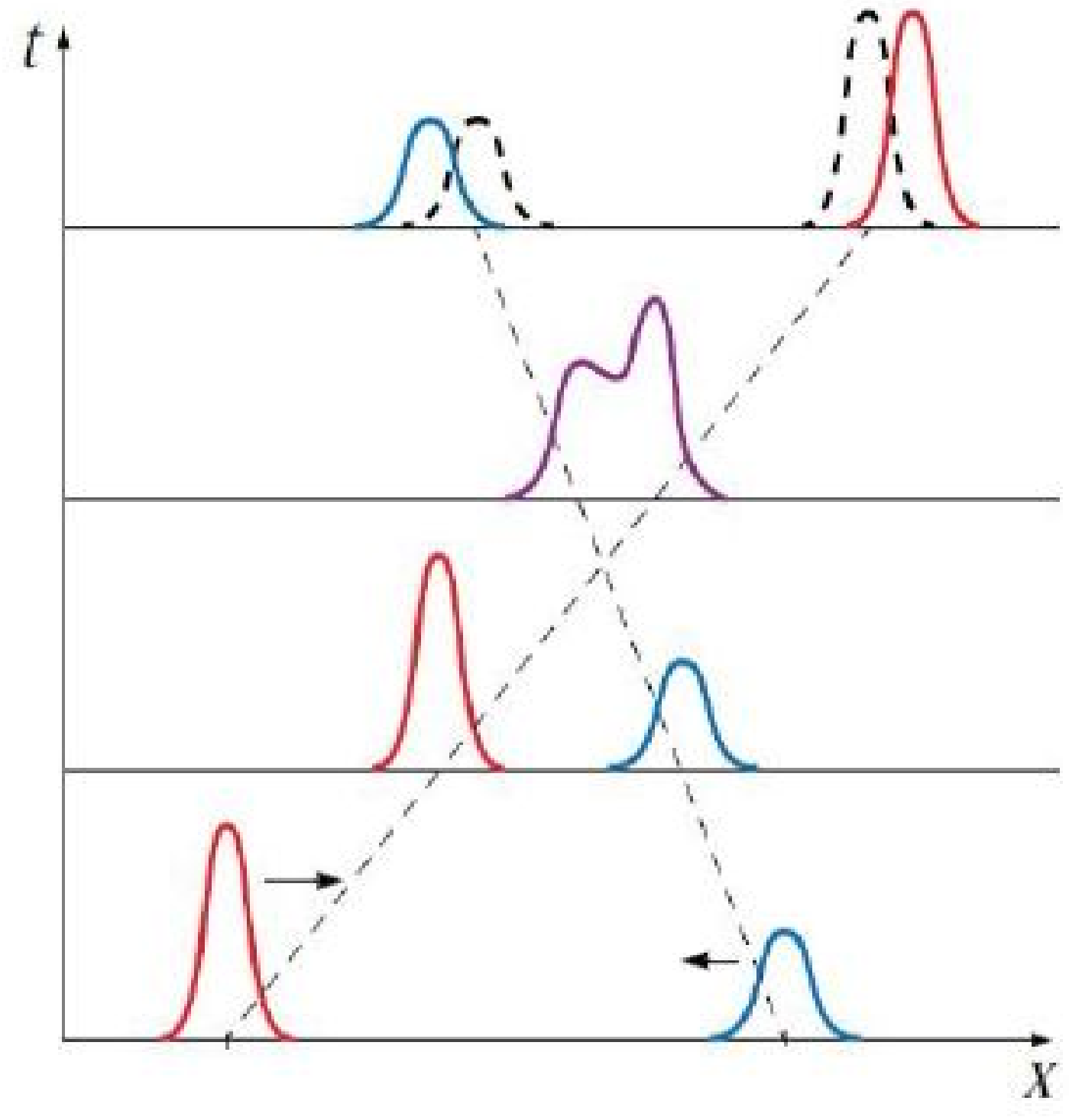}
\caption{A schematic representation of a collision between two solitary waves.}
\end{figure}

The KdV solitons are nontopological, and they exist in physical systems with weakly nonlinear and with weakly dispersive waves. When a wave impulse breaks up into several KdV solitons, they all move in the same direction (see, for example, Fig. 3). The collision of two KdV solitons Fig .8, Under certain conditions, the KdV solitons may be regarded as particles, obeying the standard laws of Newton’s mechanics. In the presence of dissipative effects (friction), the KdV solitons gradually decelerate and become smaller and longer, thus, they are “mortal.”

{\bf B. The topological solitons}

 The basic properties of a topological (kink) soliton shown in Fig. 7b can be summarized as follows [7]:

i. Its amplitude is independent of its velocity—it is constant and remains the same for zero velocity, thus the kink may be static.

ii. Its width gets narrower as its velocity increases, owing to Lorentz contraction.

iii. It has the properties of a relativistic particle.

iv. The topological kink which has a different screw sense is called an antikink.
 
Topological solitons are extremely stable. Under the influence of friction, these solitons only slow down and eventually stop and, at rest, they can live “eternally.” In an infinite system, the topological soliton can only be destroyed by moving a semi-infinite segment of the system above a potential maximum. This would require an infinite energy. However, the topological soliton can be annihilated in a collision between a soliton and an antisoliton. In an integrable system having exact soliton solutions, solitons and anti-solitons simply pass through each other with a phase shift, as all solitons do, but in a real system like the pendulum chain which has some dissipation
of energy, the soliton-antisoliton equation may destroy the nonlinear excitations. Figure .9 schematically shows a collision of a kink and an antikink in an integrable system which has soliton solutions. In integrable systems, the soliton-breather and breather-breather collisions are similar to the kink-antikink collision shown in Fig. 9.

\begin{figure}
\centering
\includegraphics[width=4 in]{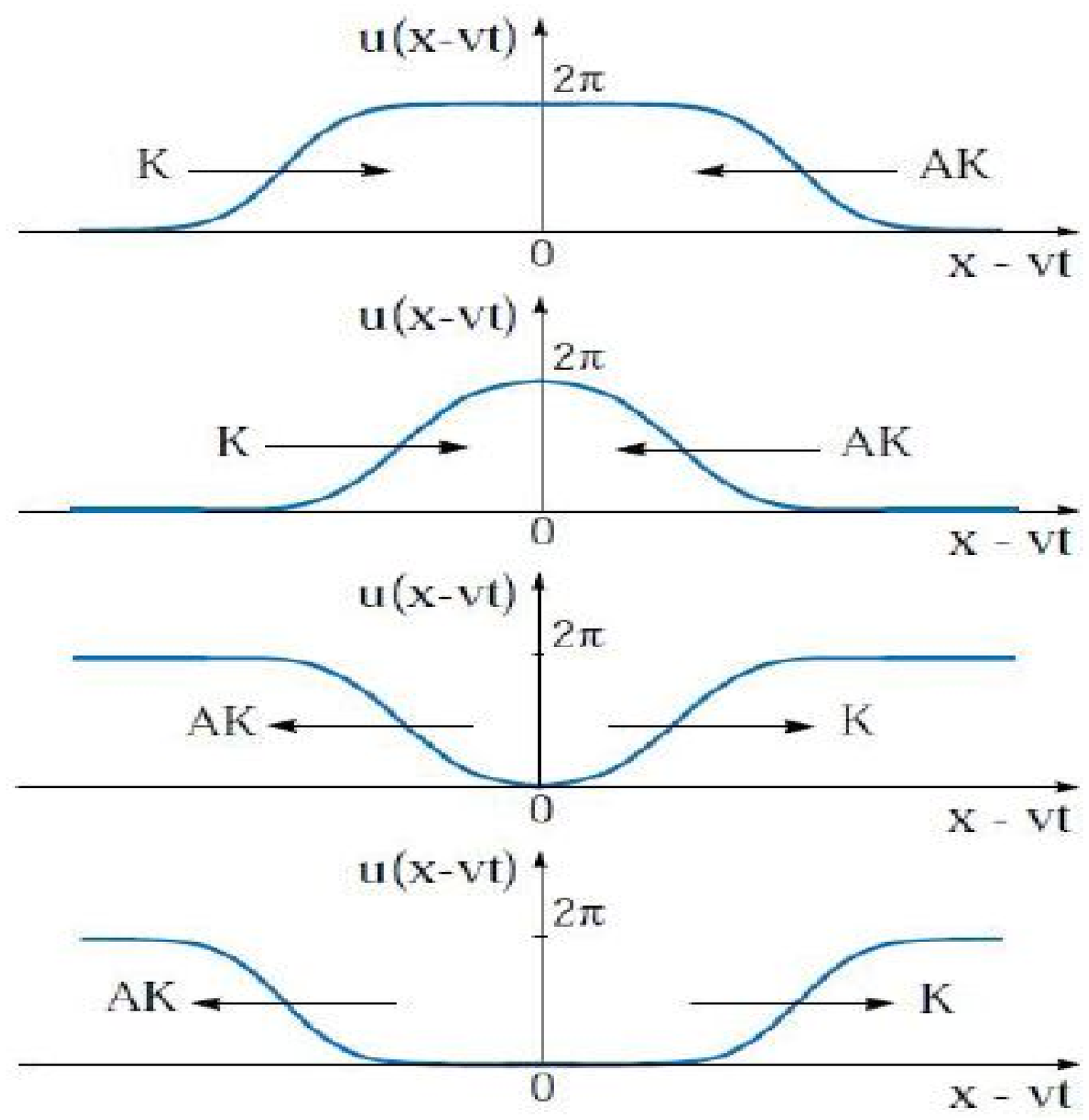}
\caption{Sketch of a collision between a kink (K) and an antikink (AK).}
\end{figure}

The sine-Gordon equation has almost become ubiquitous in the theory of condensed matter, since it is the simplest nonlinear wave equation in a periodic medium.

{\bf C. The envelope solitons}

The NLS equation is called the nonlinear Schr¨odinger equation because it is formally similar to the Schr¨odinger equation of quantum mechanics

\begin{eqnarray}
(i\bar h \frac{\partial}{\partial t}+\frac{\bar h^2}{2m}\frac{\partial^2}{\partial x^2})\psi(x,t)=0
\end{eqnarray}

where U is the potential, and $\psi (x, t)$  is the wave function.
 The NLS equation describes self-focusing phenomena in nonlinear optics, one-dimensional self-modulation of monochromatic waves, in nonlinear plasma etc. In the NLS equation, the potential U is replaced by $ |u|^2$  which brings into the system self-interaction. The second term of the NLS equation is responsible for the dispersion, and the third one for the nonlinearity. A solution of the NLS equation is schematically shown in Fig. 7c. The shape of the enveloping curve (the dashed line in Fig..7c) is given by

\begin{eqnarray}
u(x,t)=u_0\times sech((x-vt)/\ell)
\end{eqnarray}

where  $2\ell$  determines the width of the soliton. Its amplitude $ u_0 $ depends on $\ell $ , but the velocity $ \nu $ is independent of the amplitude, distinct from the KdV soliton. The shapes of the envelope and KdV solitons are also different: the KdV soliton has a sech2 shape. Thus, the envelope soliton has a slightly wider shape. However, other properties of the envelope solitons are similar to the KdV solitons, thus, they are “mortal” and can be regarded as particles. The interaction between two envelope solitons is similar to the interactions between two KdV solitons (or two topological solitons). 

In the envelope soliton, the stable groups have normally from 14 to 20 humps under the envelope, the central one being the highest one. The groups with more humps are unstable and break up into smaller ones. The waves inside the envelope move with a velocity that differs from the velocity of the soliton, thus, the envelope soliton has an internal dynamics. The relative motion of the envelope and carrier wave is responsible for the internal dynamics of the NLS soliton. 
The NLS equation is inseparable part of nonlinear optics where the envelope solitons are usually called dark and bright solitons, and became quasi-three-dimensional.

\end{document}